\documentclass[runningheads, anonymous]{llncs}
\usepackage[T1]{fontenc}
\usepackage{graphicx}
\usepackage{textcomp}
\usepackage{stfloats}
\usepackage{url}
\usepackage{verbatim}
\usepackage{cite}

\usepackage{xcolor}
\usepackage{fancyhdr}
\usepackage{booktabs}
\usepackage{svg}
\usepackage{enumitem}
\usepackage{multirow}
\usepackage{mathrsfs}
\usepackage{amsmath}
\usepackage{makecell}
\usepackage{amssymb}
\usepackage{caption}
\usepackage{float} 
\usepackage{textcomp}
\usepackage{graphicx}
\usepackage{bbding}
\newcommand\blfootnote[1]{%
  \begingroup
  \renewcommand\thefootnote{}%
  \footnote{#1}%
  \addtocounter{footnote}{-1}%
  \endgroup
}

\author{
Chongjun Xia\inst{1} \and 
Yanchun Peng\inst{2} \and
Xianzhi Wang\inst{1}\Envelope
}
\authorrunning{C. Xia et al.}

\institute{University of Technology Sydney, Sydney, Australia
\email{Chongjun.Xia@student.uts.edu.au}, \email{XIANZHI.WANG@uts.edu.au} \and
University of Shanghai for Science and Technology, Shanghai, China\\
\email{pyc6669@gmail.com}
}
\begin{document}

\title{LLM-Enhanced Reinforcement Learning for Long-Term User Satisfaction in Interactive Recommendation}
\sloppy
\maketitle              % typeset the header of the contribution
\blfootnote{C. Xia and Y. Peng---Contribute equally to this paper.}

\begin{abstract}
Interactive recommender systems can dynamically adapt to user feedback, but often suffer from content homogeneity and filter bubble effects due to overfitting short-term user preferences. While recent efforts aim to improve content diversity, they predominantly operate in static or one-shot settings, neglecting the long-term evolution of user interests. Reinforcement learning provides a principled framework for optimizing long-term user satisfaction by modeling sequential decision-making processes. However, its application in recommendation is hindered by sparse, long-tailed user-item interactions and limited semantic planning capabilities. In this work, we propose \underline{L}LM-\underline{E}nhanced \underline{R}einforcement \underline{L}earning (LERL), a novel hierarchical recommendation framework that integrates the semantic planning power of LLM with the fine-grained adaptability of RL. LERL consists of a high-level LLM-based planner that selects semantically diverse content categories, and a low-level RL policy that recommends personalized items within the selected semantic space. This hierarchical design narrows the action space, enhances planning efficiency, and mitigates overexposure to redundant content. Extensive experiments on real-world datasets demonstrate that LERL significantly improves long-term user satisfaction when compared with state-of-the-art baselines. The implementation of LERL is available at \url{https://github.com/1163710212/LERL}.
\end{abstract}

\section{Introduction}
Interactive recommender systems (IRS) enable real-time personalization through continuous adaptation to user feedback and have become integral to modern online platforms~\cite{liang2023enabling, ma2023sgnr,chongjun2025}.
However, they often overly rely on users’ short-term preferences due to overfitting, causing repeated exposure to semantically similar items or content homogeneity~\cite{gao2023cirs}. This exacerbates the well-known filter bubble effect, i.e., users are only recommended content that aligns with their past behavior and viewpoints, diminishing user satisfaction in the long run.

Recent studies have increasingly focused on promoting content diversity to mitigate filter bubble effects, such as belief-aware re-ranking~\cite{aridor2020deconstructing}, representation learning that is sensitive to echo chambers~\cite{tommasel2021want}, and diversity-aware candidate filtering~\cite{xu2020neural}.
Although these methods enhance one-shot diversity, they typically operate in static or one-shot settings and fail to consider the dynamic evolution of user preferences~\cite{gao2023cirs}. As a result, users may face content redundancy over time, reducing content diversity and users' long-term satisfaction. Recognizing this limitation, emerging research has shifted toward long-term recommendation, which aims to strategically optimize user satisfaction over extended interaction trajectories. Reinforcement learning (RL) provides a natural paradigm for this setting, given its capability to explicitly model sequential decision-making and optimize cumulative rewards. Recent RL-based approaches offer a promising path to filter bubble mitigation by modeling over-exposure dynamics~\cite{gao2023cirs}, incorporating diversity-driven intrinsic rewards~\cite{shi2023relieving}, and applying trajectory-level diversity regularization~\cite{gao2023alleviating}.
Despite active efforts, applying RL in recommender systems remains challenging due to the inherent sparsity and long-tail distribution of user-item interactions. Those characteristics hinder the recommender’s ability to generalize and plan effectively, leading to suboptimal results~\cite{rein2024}.

Large Language Models (LLMs) emerge as a powerful tool for high-level reasoning and long-term planning, a capability derived from their pretraining on massive, diverse corpora~\cite{wang2024survey}. Their strong generalization and multi-step inference capabilities make them well-suited for optimizing diversity-aware, long-term objectives. Nevertheless, LLMs often struggle to ground these abstract plans into fine-grained actions, particularly in recommender systems where the item space is extremely large and complex~\cite{wei-etal-2025-plangenllms}.

To address the above limitations with RL and LLMs, we propose \underline{L}LM-\underline{E}nhanced \underline{R}einforcement \underline{L}earning (\textbf{LERL}), a novel hierarchical recommendation framework that integrates LLM-based semantic planning with RL-based policy optimization to leverage the complementary strengths of both approaches. LERL decomposes the recommendation process into two levels of activities: 1) \textbf{High-level semantic planner:} powered by LLM, this component selects a subset of content categories based on the user's interaction history. It prioritizes semantic diversity and mitigates overexposure by deprioritizing recently saturated content types.
2) \textbf{Low-level policy learner:} trained via RL, this component generates fine-grained item-level recommendations within the candidate space constrained by the high-level planner, thereby tailoring results to individual preferences.
This hierarchical design enables the framework to leverage the structured world knowledge and long-horizon planning capabilities of LLM while retaining the adaptability and online learning strengths of RL.
By semantically narrowing the action space and explicitly promoting diversity, LERL enhances recommendation stability and fosters long-term user satisfaction.

Our contributions in the paper are summarized as follows:
\begin{itemize}
    \item We highlight the limitations of existing diversity-enhancing approaches in capturing long-term user satisfaction, particularly in scenarios of sparse and long-tailed user-item interactions.
    \item We propose a novel hierarchical recommendation framework (LERL) that integrates LLM-based planning with RL-based fine-grained policy learning to jointly optimize accuracy and diversity over time.
    \item Our extensive experiments on real-world datasets demonstrate that LERL outperforms state-of-the-art baselines in promoting long-term user satisfaction.
\end{itemize}
\section{Related Work}
\subsection{RL for Recommendation} 
Reinforcement learning (RL)-based approaches formulate interactive recommendation as a Markov Decision Process (MDP). RL-based approaches have the unique advantage of optimizing long-term cumulative user satisfaction by explicitly modeling the sequential and dynamic nature of user–item interactions. Recent advances have witnessed a surge of interest in applying RL to interactive recommender systems~\cite{liu2023redrl, gao2023alleviating}. For example, KCRL~\cite{nie2023knowledge} captures causal group effects and enhances reward estimation and state representation through a self-constructed user–item knowledge graph; Li \textit{et al.}~\cite{li2022diverse} leverage an LSTM to jointly encode cross-view interaction histories and visual states for diversified interactive recommendation; HER4IF~\cite{xia2024hierarchical} models users’ real-time preferences for popular content while enforcing fairness constraints to ensure long-term exposure fairness across item groups. Our method differs from prior RL-based approaches in introducing a hierarchical framework that leverages LLMs for high-level semantic planning---this enables more informed, diversity-aware decision-making across long horizons, complementing the fine-grained adaptability of RL in dynamic environments.

\subsection{LLM for Recommendation}
Large Language Model (LLM)-based recommendations have recently gained increasing attention, given the emergent capabilities of models such as GPT-4~\cite{achiam2023gpt}, and Llama-3~\cite{dubey2024llama} in reasoning, personalization, and interactive decision-making. Most existing LLM-powered recommendation paradigms focus on leveraging in-context learning~\cite{wei2024llmrec,xi2024towards} or fine-tuning~\cite{zhang2025recommendation} to generate immediate responses, such as item ranking or explanation generation. These methods excel at zero-shot adaptability and natural language interaction but often fall short in capturing long-term user satisfaction and behavioral dynamics. In contrast, our approach decomposes the recommendation process into high-level LLM-based semantic planning and low-level RL-based sequential optimization. The LLM module promotes category-level diversity by filtering overexposed content categories, while the RL agent further tailors item-level recommendations to user preferences. This new design enables the system to better sustain user engagement and long-term satisfaction.

\subsection{Filter Bubbles in Recommendation} 
The filter bubble problem refers to the tendency of recommender systems to repeatedly surface homogeneous content, which limits users' exposure to diverse information due to overfitting user history~\cite{areeb2023filter, tahir2024mitigating}. Prolonged exposure to such homogenized recommendations can lead to cognitive fatigue, reduced novelty perception, and ultimately, a decline in long-term user satisfaction~\cite{areeb2023filter}.
Therefore, recent work has introduced diversity-aware recommendation strategies to address filter bubbles and improve long-term user engagement.
Typically, TD-VAE-CF~\cite{gao2022mitigating} and DOR~\cite{du2021diversity} enhance content diversity through static re-ranking or diversity-aware embeddings. These approaches, however, lack sequential modeling and cannot optimize long-term outcomes.
Reinforcement learning (RL), in contrast, naturally supports long-horizon optimization in interactive recommendation, and recent RL-based methods have incorporated diversity signals to mitigate filter bubbles~\cite{li2023breaking, gao2023alleviating}.
For example, CIRS~\cite{gao2023cirs} penalizes item overexposure via satiation modeling, and DNaIR~\cite{shi2023relieving} leverages diversity-driven intrinsic rewards to promote exploration. However, these methods struggle with sparse and long-tail data, leading to suboptimal recommendations. Our proposed LERL framework complements the above efforts by incorporating an LLM-based semantic planner that directs exploration toward semantically diverse content regions. Through its hierarchical structure, LERL enables the system to make informed, diversity-aware planning decisions while preserving the adaptability and feedback-driven learning advantages of RL.

\section{Preliminary}

\subsection{Problem Formulation}
We consider an interactive recommendation scenario where an agent engages with a user over a sequence of decision steps. At each time step \( t \), the agent observes a state \( s_t \) derived from the user’s interaction history \( H_t \), and selects a recommendation list \( a_t \subseteq \mathcal{I} \); in response, the environment returns a reward \( r_t \) and transitions to the next state \( s_{t+1} \). The goal is to learn a policy \( \pi \) that maximizes long-term user satisfaction by balancing recommendation accuracy and content diversity. To this end, we adopt a hierarchical decision-making framework composed of two modules:
\begin{itemize}
\setlength{\itemsep}{0.3em} % Sets the spacing after each item to 1em
    \item \textbf{High-Level Semantic Planner:} A LLM-based agent analyzes the user’s interaction trajectory and selects a subset of candidate categories \( c_t \subseteq \mathcal{C} \), discouraging overexposed content types.
    \item \textbf{Low-Level Policy Learner:} A RL-based agent selects the recommendation list \( a_t \subseteq \mathcal{I}_{c_t} \) to maximize the expected long-term reward, where \( \mathcal{I}_{c_t} \) denotes the set of items in the selected categories $c_t$.
\end{itemize}

The interaction trajectory can be represented as:
\begin{equation}
\mathcal{T}_{1:N} = \{(s_1, c_1, a_1, r_1), \ldots, (s_N, c_N, a_N, r_N)\},
\end{equation}
where \( c_t \) is the category set selected by the LLM at step \( t \), and \( a_t \subseteq \mathcal{I}_{c_t} \) is the item chosen by the RL agent. The optimization objective is defined as:
\begin{equation}
\max_{\pi} \ \mathbb{E}_{\pi} \left[ \sum_{t=1}^{N} \gamma^{t-1} r_t \right],
\end{equation}
where \( \gamma \) is the discount factor that balances immediate and long-term rewards.

\subsection{Simulated Environment}
As collecting online user feedback can be financially burdensome, following prior work~\cite{zhao2023kuaisim}, we construct a simulated offline environment based on logged user-item interaction datasets. The simulator emulates real-time user feedback and session-level behaviors, including item-level engagement and user dropout. This enables safe and reproducible evaluation of interactive recommendation policies under realistic dynamics.
More details about the simulated environment can be found in Section~\ref{env}.

\begin{figure*}[!t]
    \centering
    \includegraphics[width=0.92\linewidth]{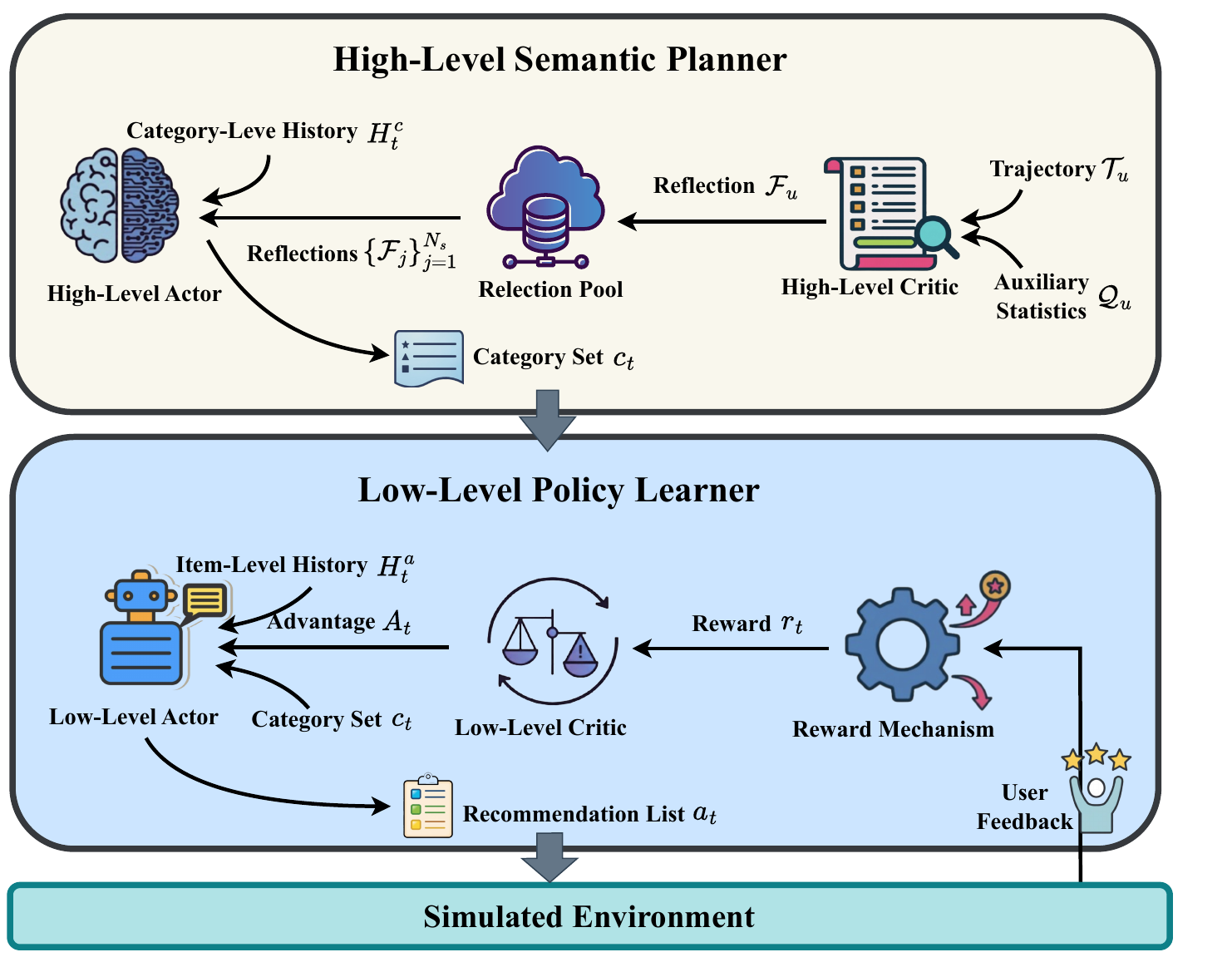}
    \caption{Overview of the LERL framework.}
    \label{fig:overal_architecture}
\end{figure*}
\section{Methodology}
We propose a hierarchical agent framework, LERL, which integrates two levels of decision-making for interactive recommendation: a high-level semantic planner (HSP) powered by an LLM and a low-level policy learner (LPL) optimized via RL. The LLM-based agent generates category-level constraints by determining the candidate item categories, while the RL-based agent performs fine-grained item selection within these constraints. This hierarchical design effectively balances category-level diversity and user preference alignment, promoting long-term user satisfaction. The overall architecture of LERL is shown in Fig.~\ref{fig:overal_architecture}.

\subsection{High-Level Semantic Planner}
The HSP functions as a macro-level decision-maker, exerting strategic control over category-level exposure. By leveraging a pretrained LLM, the HSP performs semantic planning over user intent and interaction context at the category granularity, effectively balancing recommendation accuracy and content diversity.

\subsubsection{High-Level Actor}
At each decision step $t$, the high-level actor selects a category set $c_t \subseteq \mathcal{C}$, where $\mathcal{C}$ denotes the set of all candidate item categories. This decision is based on the user’s category-level interaction history and a set of reflections that summarize the complete interaction trajectories of other users. Formally, the user’s interaction history up to step $t$ is denoted as:
\begin{equation}
    H_t^c = \left\{(c_i, r_i)\right\}_{i=1}^{t},
\end{equation}
where $c_i$ is the selected category set at step $i$, and $r_i$ is the user's immediate satisfaction signal (e.g., click, rating).

To enhance decision-making, we maintain a \textit{reflection pool} \( \mathcal{R} = \{(\mathcal{F}_u, S_u)\}_{u=1}^{N_r} \), where each \( \mathcal{F}_u \) denotes the category-level reflection (generated by the high-level critic) over the full interaction trajectory of a past user \( u \). The associated score \( S_u = \sum_{t=1}^{T_u} r_t \) represents the cumulative reward, indicating their long-term satisfaction. \( T_u \) denotes the length of the interaction session for user \( u \).
Given the limited context length of LLM, it is infeasible to incorporate the entire reflection pool into a single prompt. Therefore, we design a sampling mechanism based on users' cumulative rewards to prioritize more informative reflections. Specifically, the sampling distribution \( P(u) \) over the reflection pool is defined as follows:
\begin{equation}
P(u) = \frac{\exp(\alpha S_u)}{\sum_{v=1}^{N_r} \exp(\alpha S_v)},
\end{equation}
where \( \alpha \) is a temperature parameter controlling the sharpness of the distribution.

Then, a set of \( N_s \) reflections \( \{ \mathcal{F}_j \}_{j=1}^{N_s} \) is sampled from the full reflection pool \( \mathcal{R} \) according to a user-specific distribution \( P(u) \). The selected reflections serve as high-level semantic guidance for planning. Note that in the early stages of interaction, when no user sessions have been completed, the reflection pool is empty, and the set \( \{ \mathcal{F}_j \}_{j=1}^{N_s} \) is substituted with empty placeholders. 

\begin{figure}[!t]
  {
    \centering
    \includegraphics[width=0.9\textwidth]{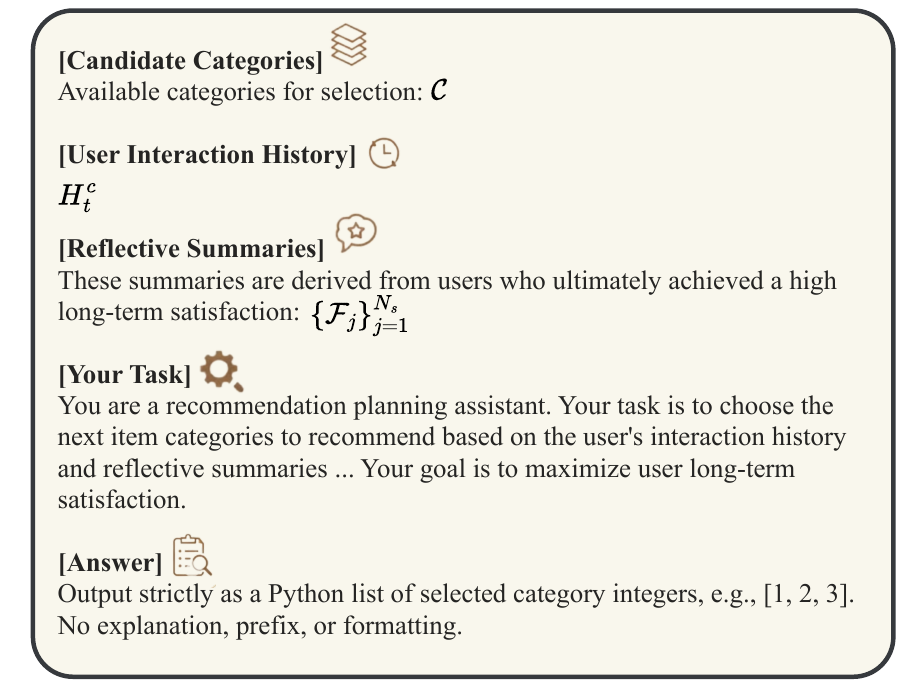}
    \caption{Prompt design for the high-level actor, which selects category set by integrating category-level interaction history and sampled reflections.}
    \label{fig:prompt_h_actor}
  }
\end{figure}

Finally, the category set is selected via prompt-based querying of the LLM:
\begin{equation}
c_t = \text{LLM}\left(\text{Prompt}_\text{a}(\mathcal{C}, H_t^c, \{\mathcal{F}_j\}_{j=1}^{N_s})\right),
\end{equation}
where \( \text{Prompt}_\text{a} \) is a structured natural language template that integrates candidate categories \( \mathcal{C} \), category-level interaction history \( H_t^c \), and the sampled reflections $\{\mathcal{F}_j\}_{j=1}^{N_s}$ (an example is shown in Fig.~\ref{fig:prompt_h_actor}). This design enables the LLM to perform high-level semantic planning, striking a balance between recommendation relevance and content diversity.

\subsubsection{High-Level Critic}
\begin{figure}[!t]
  {
    \centering
    \includegraphics[width=0.9\textwidth]{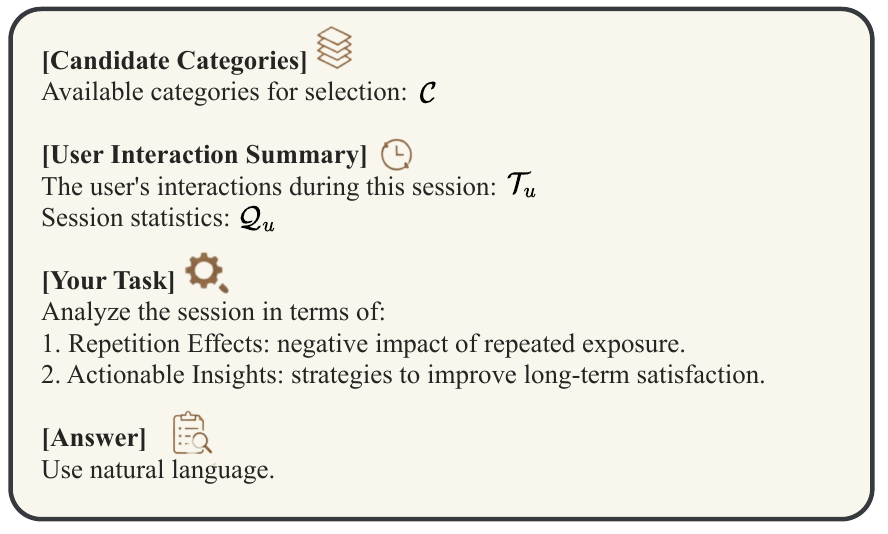}
    \caption{Prompt design for the high-level critic, which generates the reflection from category-level interactions and auxiliary statistics in a user's session.}
    \label{fig:prompt_h_critic}
  }
\end{figure}

We propose a \textit{reflective language-based critic} to support self-improvement in high-level decision-making. Unlike traditional scalar reward signals, this module outputs a textual reflection $\mathcal{F}_u$ at the end of user $u$'s session, encapsulating actionable insights that contribute to enhancing long-term user satisfaction. The reflection is generated as follows:
\begin{equation}
\mathcal{F}_u = \text{LLM}\left(\text{Prompt}_\text{v}(\mathcal{C}, \mathcal{T}_u, \mathcal{Q}_u)\right),
\end{equation}
where $\text{Prompt}_\text{v}$ is a structured natural language template (an example is shown in Fig.~\ref{fig:prompt_h_critic}). $\mathcal{C}$ denotes the set of candidate categories, and $\mathcal{Q}_u$ represents auxiliary statistics, e.g., \textit{interaction length} and \textit{cumulative reward}. $\mathcal{T}_u = \{(c_i, r_i)\}_{i=1}^T$ denotes the complete category-level interaction trajectory. These reflections serve as high-level semantic critiques, guiding the model toward strategies that promote long-term user engagement. The generated reflections are stored in the reelection pool and reused in future prompts to inform and condition the high-level actor’s decision-making.

\subsection{Low-Level Policy Learner}
The LPL focuses on item selection. It adheres to high-level constraints $c_t$ while optimizing for user satisfaction signals, e.g., clicks or ratings. This design allows the LPL to specialize in fine-grained ranking quality while delegating strategic control to the HSP.

\subsubsection{Low-Level Actor} 
The low-level actor generates a recommendation list $a_t$ conditioned on the low-level state $s_t$ and the high-level constraint $c_t$. The low-level state \( s_t \) encodes the user’s item-level interaction history \( H_t^a = \left\{(a_i, r_i)\right\}_{i=1}^{t}\), where $r_i$ is the user's immediate satisfaction signal, e.g., a click or rating.

\textbf{First}, we adopt a Transformer-based encoder to capture the sequential dependencies among recent interactions. Specifically, each interaction is passed through an embedding layer to produce a sequence of hidden representations \( V_t=\{v_{1}, v_{2}, \dots, v_t\} \), which is fed into the Transformer to derive user preference features:
\begin{equation}
    e_t^p = \text{Transformer}(v_{1}, v_{2}, \dots, v_t)
\end{equation}

The preference representation \( e_t^p \) is then passed through a multi-layer perceptron (MLP) to parameterize a Gaussian distribution with mean \( \mu_t \) and standard deviation \( \sigma_t \). This stochastic modeling facilitates continuous exploration in the policy space and helps prevent premature convergence to suboptimal solutions.
To capture the user's latent intent in the embedding space, a virtual item embedding \( p_t \) is sampled from the above distribution as follows:
\begin{equation}
p_t \sim \mathcal{N}(\mu_t, \sigma_t^2).
\end{equation}

The similarity score between each candidate item embedding \( i_j \) and \( p_t \) is computed via dot product:
\begin{equation}
a^{\text{sim}}_j = \langle p_t, i_j \rangle, \quad \forall j \in \{1, \dots, |\mathcal{I}|\}
\end{equation}

\textbf{Second}, we apply category-level constraints guided by the high-level action \( c_t \) as a soft filter over candidate items. Specifically, we calculate the corresponding category mask as:
\begin{equation}
\label{eq:mask}
a^{\text{mask}} = W \cdot c_t,
\end{equation}
where \( W \in \{0,1\}^{|\mathcal{I}| \times |\mathcal{C}|} \) is a binary item-category mapping matrix. Each row of \( W \) corresponds to an item \( i \in \mathcal{I} \), and each column corresponds to a category \( c \in \mathcal{C} \). An entry of 1 indicates the item belongs to the corresponding category.

\textbf{Last}, we apply the category mask to the similarity scores to compute the item-level recommendation scores (see Eq.~\eqref{recscore}) and select the top-\( k \) items with the highest values in \( a^{\text{score}} \) to form the recommendation list $a_t$.
\begin{equation}
\label{recscore}
a^{\text{score}} = a^{\text{sim}} \odot a^{\text{mask}},
\end{equation}
where \( \odot \) denotes element-wise multiplication.

\subsubsection{Low-Level Critic}

We adopt a value estimator in the form of a state-value function $V_{\varphi}(s_t)$ to guide the learning of the low-level actor. This value function is approximated using a multi-layer perceptron:
\begin{equation}
    V_{\varphi}(s_t) = \text{MLP}_v(s_t),
\end{equation}
where $s_t$ denotes the low-level state at time step $t$.

The low-level agent is optimized using the proximal policy optimization (PPO) algorithm~\cite{schulman2017proximal}, in light of its strong empirical performance and stability in policy gradient updates.
The critic estimates the expected return from state $s_t$, and is trained by minimizing the temporal difference error:
\begin{equation}
\label{eq:loss_value_h}
\mathcal{L}_v = \mathbb{E}_t \left[ \left( V_{\varphi}(s_t) - R_t \right)^2 \right], \quad
R_t = r_t + \gamma V_{\varphi'}(s_{t+1}),
\end{equation}
where \( \varphi \) and \( \varphi' \) denote the parameters of the online and target critic networks, respectively; $r_i$ is the user's immediate satisfaction signal (e.g., click or rating); and \( \gamma \in [0,1] \) is the discount factor.

The corresponding actor is trained by maximizing the clipped surrogate objective of PPO, which balances learning efficiency and stability:
\begin{equation}
\begin{split}
\mathcal{L}_{a} = -\mathbb{E}_t\left[
\min \left(
\frac{\pi_{\theta}\left(a_t \mid s_t\right)}{\pi_{\theta_{\text{old}}}\left(a_t \mid s_t\right)} \hat{A}_t,\ 
\text{clip}\left(
\frac{\pi_{\theta}\left(a_t \mid s_t\right)}{\pi_{\theta_{\text{old}}}\left(a_t \mid s_t\right)},
1 - \epsilon,\ 1 + \epsilon
\right) \hat{A}_t
\right)
\right],
\end{split}
\end{equation}
where $\theta$ and $\theta_{\text{old}}$ are the parameters of the main and old policy networks, respectively. $\hat{A}_t$ is the estimated advantage function, and $\epsilon$ is a hyperparameter that bounds the policy update ratio to prevent destructive updates. The $\text{clip}(x, a, b)$ function restricts $x$ within the interval $[a, b]$. For further algorithmic details and theoretical justification, please refer to the original paper~\cite{schulman2017proximal}.

\section{Experiments}
This section reports our empirical evaluation of the proposed method in interactive recommendation settings. Specifically, we designed experiments to address the following questions:

\begin{itemize}[label=\small$\bullet$, leftmargin=1.2em]
  \item \textbf{RQ1}: How does LERL compare to state-of-the-art RL-based methods for interactive recommendation?
  \item \textbf{RQ2}: How well does LERL mitigate the filter bubble effect in recommendation trajectories?
  \item \textbf{RQ3}: How do core modules of LERL contribute to the overall recommendation performance?
\end{itemize}

\subsection{Experiment Setup}
%This section reports our experimental settings, including simulation environment, evaluation metrics, baselines, and implementation details.

\subsubsection{Simulation Environment}
\label{env}

Following prior work~\cite{yu2024easyrl4rec, shi2023relieving}, we conducted simulation-based experiments in the KuaiSim~\cite{zhao2023kuaisim}, as direct online evaluation can be prohibitively costly. KuaiSim is a comprehensive user simulator that supports request-level, whole-session, and cross-session recommendation tasks. In this work, we leveraged KuaiSim to conduct whole-session recommendation experiments. Within this framework, the environment provides the agent with sequential interaction signals, including predicted rewards and dynamically evolving user states based on historical behaviors. This simulation setup enables a systematic evaluation of models in their ability to mitigate filter bubble effects and improve long-term user satisfaction. Simulation environments were constructed based on two high-quality datasets: KuaiRand (pure, 15 policies)~\cite{gao2022kuairand} and KuaiRec (big matrix)~\cite{gao2022kuairec}. For both datasets, the “is\_click” field was used to represent user interest and serves as the reward signal in the simulated environment. Table~\ref{tab:dataset_statistics} shows basic statistics of the two datasets.
\begin{table}[!t]
  \centering
  \caption{Statistics of experimental datasets}
  \setlength{\tabcolsep}{3pt} % change the horizontal spacing (column separation)
  \begin{tabular}{lcccc}
  \toprule
  \textbf{Dataset} & \textbf{\#Users} & \textbf{\#Items} & \textbf{\#Interactions} &  \textbf{Density} \\ 
  \midrule
  \multirow{1}{*}{KuaiRand}  & 27,077 & 7,551  & 1,436,609  & 0.70\% \\ 
  \multirow{1}{*}{KuaiRec}  & 7,176 & 10,728  & 12,530,806  & 16.3\%\\ 
  \bottomrule
  \end{tabular}
  \label{tab:dataset_statistics}
\end{table}

Following~\cite{zhao2023kuaisim}, we set the maximum interaction length to 20, considering that real users typically engage in a limited number of interactions within a single session. As such, each user is allowed to generate up to 20 interactions per session in the simulation environment. The length of each recommendation list was set to 6. To effectively capture the impact of filter bubbles, we introduced a \textbf{diversity-aware quit mechanism} inspired by prior work~\cite{quanavlrecom2023, gao2023cirs}. Specifically, the user's maximum remaining interaction count is decremented by one whenever the recommendation agent consecutively generates recommendation lists containing items from the same category. This penalty reflects user boredom caused by excessive exposure to homogeneous content, thereby simulating the early termination of sessions under low content diversity.

\subsubsection{Evaluation Metrics}

Following prior studies~\cite{yu2024easyrl4rec, zhao2023kuaisim}, we evaluated the performance of interactive recommendation approaches using three metrics:
\begin{itemize}
    \item {Interaction length} ($\mathbf{T_{int}}$) denotes the number of interactions before the session terminates. Owing to the diversity-aware quit mechanism, a larger $\mathbf{T_{int}}$ indicates that the model provides more diverse recommendations.
    \item {Cumulative reward} ($\mathbf{R_{cum}}$) measures the total reward accumulated throughout the session, formally defined as $\mathbf{R_{cum}} = \sum_{t=1}^{\mathbf{T_{int}}} r_t$, where $r_t$ represents the reward at step $t$ (e.g., click). This metric reflects overall user satisfaction during the session.
    \item {Single-round reward} ($\mathbf{R_{sin}}$) is computed as the average reward per interaction step, i.e., $\mathbf{R_{sin}} = \frac{\mathbf{R_{cum}}}{\mathbf{T_{int}}}$. While $\mathbf{R_{sin}}$ captures per-step effectiveness in satisfying user preference, it does not necessarily correlate with long-term satisfaction.
\end{itemize}

\subsubsection{Baselines} 
To comprehensively and fairly evaluate the effectiveness of LERL, we made comparisons with representative RL-based and diversity-aware RL approaches. Specifically, we included the following baselines:
\begin{itemize}
\item
PG~\cite{williams1992simple}: a fundamental policy gradient method that directly optimizes the expected return via stochastic gradient ascent;
\item
A2C~\cite{konda1999actor}: an actor-critic framework that synchronously updates value and policy networks to stabilize learning;
\item
DDPG~\cite{lillicrap2015continuous}: a deterministic actor-critic algorithm with target networks and soft updates;
\item
TD3~\cite{fujimoto2018addressing}: an improved variant of DDPG that employs twin critics and policy smoothing to reduce overestimation bias; 
\item
PPO~\cite{schulman2017proximal}: a clipped surrogate objective method that constrains policy updates for stable and efficient training;
\item
HAC~\cite{2023hacwww}: a latent-action RL framework that infers vectorized hyper-actions to generate recommendations, with alignment and supervision for stable training;
\item
SAC4IR~\cite{shi2024maximum}: an entropy-regularized RL approach with a debiasing-aware reward model designed for interactive recommendation scenarios;
\item
DNaIR~\cite{shi2023relieving}: a diversity-aware RL framework that integrates item similarity, diversity, and quality signals into the reward function.
\end{itemize}

\subsubsection{Implementation Details}
All methods were fine-tuned in simulated environments. For PG and A2C, the discount factor was set to 0.9. For DDPG and TD3, the replay buffer size was set to 20{,}000. The PPO clipping threshold $\epsilon_{\text{clip}}$ was set to 0.8. For HAC, the hyper-action alignment coefficient $\lambda_h$ was set to 0.1. For SAC4IR, the relaxation parameter was set to 0.8. For DNaIR, the minibatch size was set to 64 with $\beta = 0.1$. For our method LERL, the reflection pool size $N_r$ was set to 200, and the number of sampled reflections $N_s$ was set to 3. The LLM used for the high-level module was Llama-3-8B~\cite{dubey2024llama}. The code is available at \url{https://github.com/1163710212/LERL}.

\begin{table*}[!t]
\centering
\caption{Results of all methods in two environments (\textbf{Bold}: Best; \underline{Underline}: Runner-up).}
\label{tab:overall_performance}
\setlength{\tabcolsep}{3pt} % change the horizontal spacing (column separation)
\begin{tabular}{@{}c|c|ccc@{}}
\toprule
\textbf{Env} & \textbf{ Method } & $\mathbf{T_{\text{int}}}$ & $\mathbf{R_{\text{sin}}}$ & $\mathbf{R_{\text{cum}}}$ \\
\midrule

\multirow{9}{*}{\quad KuaiRand \quad}
&PG &11.852 $\pm$ 0.305 &0.618 $\pm$ 0.033 &7.476 $\pm$ 0.399\\
&A2C &11.795 $\pm$ 0.373 &0.650 $\pm$ 0.033 &7.599 $\pm$ 0.507\\
&DDPG &11.390 $\pm$ 0.345 &0.702 $\pm$ 0.037 &8.117 $\pm$ 0.515\\
&TD3 &11.705 $\pm$ 0.318 &0.581 $\pm$ 0.039 &6.734 $\pm$ 0.542\\
&PPO &\underline{14.352 $\pm$ 0.608} &0.593 $\pm$ 0.039 &8.310 $\pm$ 0.560\\
&HAC &11.795 $\pm$ 0.648 &0.671 $\pm$ 0.050 &7.902 $\pm$ 0.602\\
&SAC4IR &11.895 $\pm$ 0.531 &0.650 $\pm$ 0.031 &7.751 $\pm$ 0.584\\
&DNaIR &13.500 $\pm$ 0.378 &\textbf{0.720 $\pm$ 0.028} &\underline{9.824 $\pm$ 0.540}\\
&LERL &\textbf{17.238 $\pm$ 0.783} &\underline{0.719 $\pm$ 0.038} &\textbf{12.284 $\pm$ 0.493}\\

\midrule

\multirow{9}{*}{\quad KuaiRec \quad} 
&PG &11.681 $\pm$ 0.263 &0.695 $\pm$ 0.031 &8.118 $\pm$ 0.418\\
&A2C &\underline{13.305 $\pm$ 0.600} &0.428 $\pm$ 0.033 &5.791 $\pm$ 0.366\\
&DDPG &11.552 $\pm$ 0.557 &0.575 $\pm$ 0.057 &6.574 $\pm$ 0.631\\
&TD3 &11.043 $\pm$ 0.192 &\textbf{0.800 $\pm$ 0.046} &8.771 $\pm$ 0.489\\
&PPO &11.914 $\pm$ 0.315 &0.676 $\pm$ 0.047 &7.969 $\pm$ 0.573\\
&HAC &11.000 $\pm$ 0.000 &0.546 $\pm$ 0.030 &5.953 $\pm$ 0.271\\
&SAC4IR &11.167 $\pm$ 0.194 &0.522 $\pm$ 0.036 &5.781 $\pm$ 0.516\\
&DNaIR &13.105 $\pm$ 0.381 &\underline{0.710 $\pm$ 0.032} &\underline{9.235 $\pm$ 0.694}\\
&LERL &\textbf{16.400 $\pm$ 0.474} &0.637 $\pm$ 0.032 &\textbf{10.507 $\pm$ 0.543}\\

\bottomrule
\end{tabular}
\end{table*}

\subsection{Main Comparison (RQ1)}
To assess LERL’s effectiveness in promoting long-term user satisfaction, we conducted comprehensive comparisons against state-of-the-art baselines across two datasets. Our comparison results (Table~\ref{tab:overall_performance}) reveal the following findings:
\begin{itemize}
\setlength{\itemsep}{0.3em} % Sets the spacing after each item to 1em
%[leftmargin=1em]
    \item LERL consistently achieved the best overall performance in both environments. Specifically, LERL outperformed all baselines in terms of interaction length ($\text{T}_{\text{int}}$) and cumulative reward ($\text{R}_{\text{cum}}$), demonstrating its strong ability to encourage prolonged engagement and enhance long-term user satisfaction. 
    
    \item DNaIR showed moderate improvements due to its incorporation of item diversity into the reward signal and policy design. However, the gains in $\text{T}_{\text{int}}$ and $\text{R}_{\text{cum}}$ remained limited, suggesting that static or manually designed diversity rewards are ineffective in long-horizon optimization.
    
    \item Traditional RL algorithms fall short on long-term metrics. Methods like TD3 and DDPG achieved reasonable single-round reward ($\text{R}_{\text{sin}}$), but exhibited significantly lower $\text{T}_{\text{int}}$ and $\text{R}_{\text{cum}}$, highlighting their limitations in addressing content redundancy and dynamically evolving user preferences in interactive recommendation settings.

\end{itemize}

\subsection{Case Study (RQ2)}
\begin{figure}[!t]
  {
    \centering
    \includegraphics[width=0.9\textwidth]{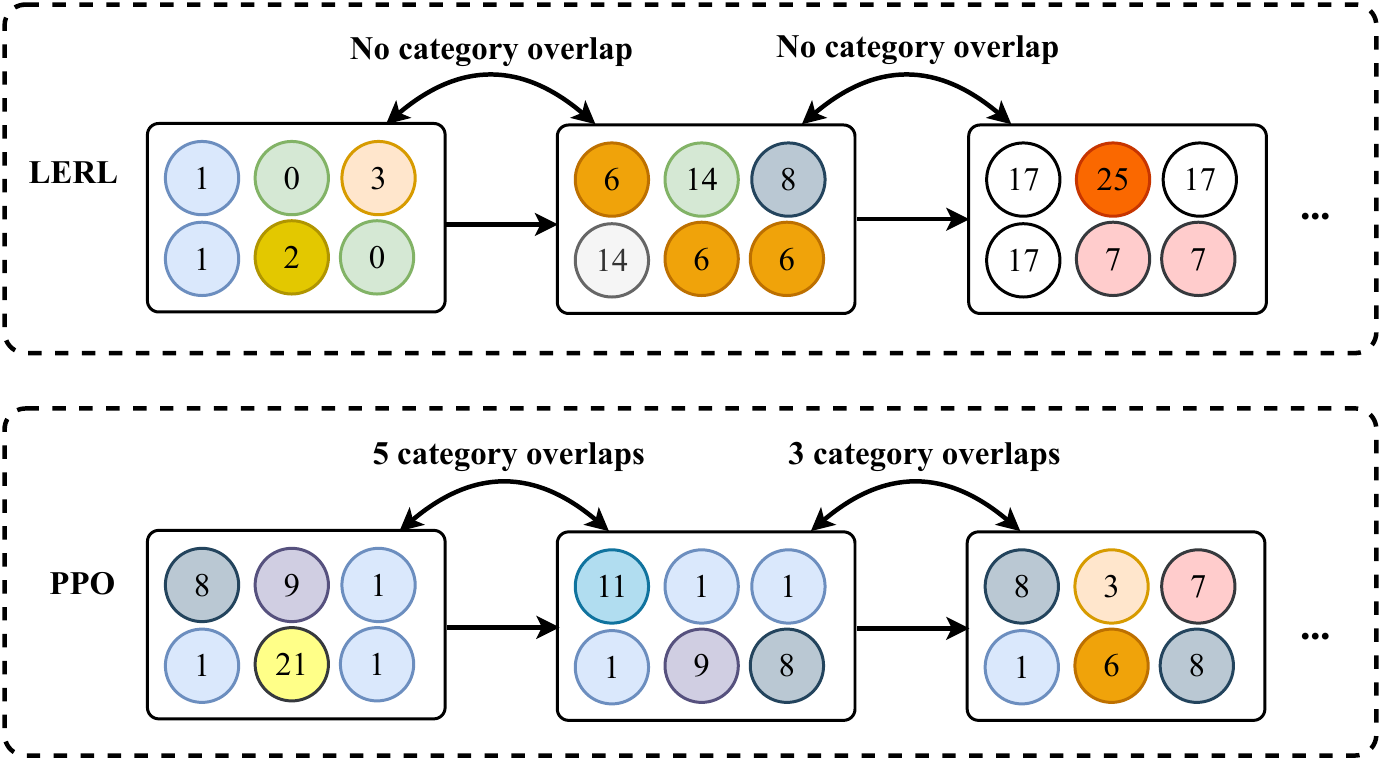}
     \caption{Recommendation trajectories of LERL and PPO.}
    \label{fig:case_study}
  }
\end{figure}

We use a case study on the KuaiRec dataset to demonstrate the effectiveness of our method in mitigating the filter bubble effect.
%, we conducted a case study using t.
Specifically, we sampled the recommendation trajectory of a user (ID: 5) and compared the results of three consecutive recommendation rounds generated by LERL and the PPO baseline (shown in Fig.~\ref{fig:case_study}), where each circle denotes an item, and the number in each circle indicates the category to which the item belongs.

%We observe that 
The recommendation lists generated by LERL exhibited no repeated item-categories throughout the three rounds; in contrast, the PPO-based lists contained a high degree of redundancy in category exposure across consecutive rounds.
This contrast highlights LERL's ability to diversify category-level exposure over time, demonstrating its superiority in alleviating content homogeneity and fostering a more engaging long-term user experience.

\subsection{Ablation Study (RQ3)}
We conducted ablation studies to assess the contribution of the two core components in the LERL framework. We denote by \textbf{w/o hsp} the variant of LERL without the high-level semantic planner, and \textbf{w/o hc} the variant that removes the high-level critic. We have focused on these two components because they represent the key innovations of LERL: the high-level planner controls diversity by generating category-level constraints, while the high-level critic offers long-term semantic reflections based on user trajectories.

\begin{figure}[!t]
  {
    \centering
    \includegraphics[width=0.96\textwidth]{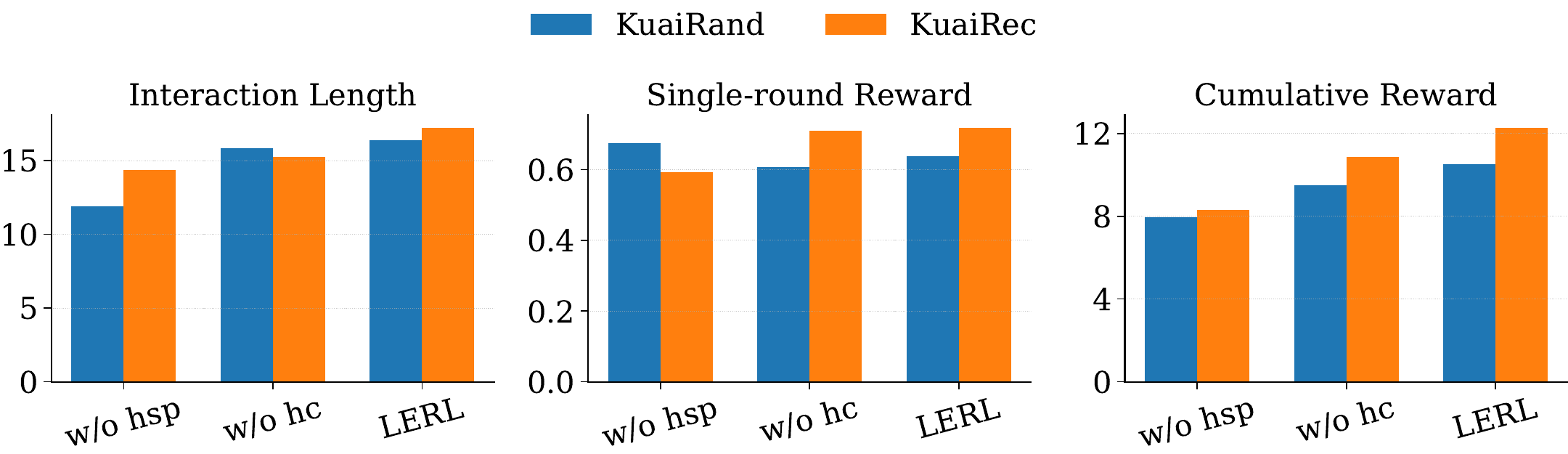}
    \caption{The results of the ablation study.}
    \label{fig:ablation_study}
  }
\end{figure}

Our ablation results (Fig.~\ref{fig:ablation_study}) show that removing either component leads to notable performance degradation in both environments. In particular, eliminating the high-level planner (\textbf{w/o hsp}) results in a substantial drop in interaction length and cumulative reward, highlighting the role of HSP in maintaining long-term engagement and satisfaction through strategic category planning. Removing the reflective critic (\textbf{w/o hc}) adversely impacts all three evaluation metrics, indicating the crucial role of semantic reflections in shaping high-level decisions.

\section{Conclusion}
This paper presents LERL, a novel hierarchical recommendation framework that integrates LLM and RL to enhance long-term user satisfaction in interactive recommender systems. LERL decomposes the recommendation process into two levels: a high-level semantic planner that selects semantically diverse content categories using LLM, and a low-level RL policy that performs fine-grained item selection under these constraints. This design effectively narrows the action space, mitigates overexposure to redundant content, and promotes semantic diversity throughout user interactions. Comprehensive experiments on simulated environments demonstrate that LERL consistently outperforms state-of-the-art baselines in long-term user satisfaction.

A promising direction for future work is to extend LERL into a multi-objective optimization framework that incorporates auxiliary goals, including fairness, diversity, and accuracy. Integrating more advanced prompting strategies and leveraging continual learning for the high-level planner also holds potential for further enhancing adaptability and robustness.

% \subsubsection{Disclosure of Interests.}The authors have no competing interests to declare that are relevant to the content of this article. 

\bibliographystyle{splncs04}
\bibliography{cite}
\end{document}